# Evolution of Magnetoresistance in the magnetic topological semimetals NdSb$_x$Te$_{2-x}$

Santosh Karki Chhetri[1], Rabindra Basnet[1,2], Krishna Pandey[3], Gokul Acharya[1], Sumaya Rahman[3], Md Rafique Un Nabi[1], Dinesh Upreti[1], Hugh O.H. Churchill[1,3,4,5], Jin Hu[1,3,4,5*]

[1]Department of Physics, University of Arkansas, Fayetteville, AR 72701, USA

[2]Department of Physics, Morgan State University, Baltimore, MD 21251, USA

[3]Materials Science and Engineering Program, Institute for Nanoscience and Engineering, University of Arkansas, Fayetteville, AR 72701, USA

[4]MonArk NSF Quantum Foundry, University of Arkansas, Fayetteville, Arkansas, 72701 USA

[5]Smart Ferroic Materials Center, University of Arkansas, Fayetteville, Arkansas, 72701 USA

# ABSTRACT

Magnetic topological semimetals $Ln$SbTe ($Ln$ = lanthanide elements) provide a platform to study the interplay of structure, magnetism, topology, and electron correlations. Varying Sb and Te compositions in $Ln$Sb$_x$Te$_{2-x}$ can effectively control the electronic, magnetic, and transport properties. Here, we report the evolution of transport properties with Sb and Te contents in NdSb$_x$Te$_{2-x}$, ($0 < x < 1$). Our work reveals nonmonotonic evolution in magnetoresistance with varying composition stoichiometry. Specifically, reducing Sb content $x$ leads to strong negative magnetoresistance up to 99.9%. Such a strong magnetoresistance, which is likely attributed to the interplay between structure, magnetism, and electronic bands, establishes this material as a promising platform for investigating topological semimetal for future device applications.

*Contact author: jinhu@uark.edu

Topological semimetals (TSMs) have emerged as a pivotal research area, offering profound insights into novel electronic phenomena and promising advancements in material science [1–4]. In their electronic structures, TSMs such as Dirac or Weyl semimetals exhibit symmetry-protected linearly dispersed Dirac or Weyl cones, which host relativistic electrons that lead to exotic properties including surface Fermi arcs [5–7], ultrahigh mobility [8–10], chiral anomaly [11,12], and large magnetoresistance [8,10,12]. Among various TSMs, magnetic TSMs such as $Co_3Sn_2S_2$ [13,14], FeSn [15], $Fe_3GeTe_2$ [16], and $Mn_3(Ge/Sn)$ [17–20] have attracted significant interest because of the presence of magnetism and its interactions with the non-trivial band topology, which leads to phenomena such as the giant anomalous Hall effect [13,21,22], anomalous Nernst effect [21,23], and spin-polarized surface states [24–26]. Usually, magnetism in magnetic TSMs originates from either $3d$ or $4f$ electrons. *Ln*SbTe (*Ln* = Rare earth elements) represents such compounds where magnetism is attributed to the $4f$ electrons of rare earth elements [27–40]. The stoichiometric *Ln*SbTe is isostructural to non-magnetic ZrSi(S/Se/Te) [41–44], which are nodal-line semimetals exhibiting linear band crossings along one-dimensional lines or loops in momentum space. These materials display a layered PbFCl-type crystal structure (space group *P*4/*nmm*) with Si square nets, hosting both Dirac nodal-line and Dirac nodal point states protected by $C_{2v}$ and non-symmorphic symmetries respectively [41,42], and displaying exotic properties such as electronic correlation enhancement [45], pressure-induced topological phase transition [46,47], and unconventional symmetry reduction-induced surface state [48,49]. With inducing $4f$ magnetism, *Ln*SbTe exhibit antiferromagnetic (AFM) ground states for various *Ln* such as Ce [29,39], Nd [27,37,50], Sm [31,51], Gd [28,52], Tb [34,35], Dy [36,37], Ho [30,38], and Er [37], with the exceptions of LaSbTe which is non-magnetic [53,54] and PrSbTe which does not display well-defined long-range order down to 2 K [32,33,40]. Under magnetic field, metamagnetic transitions have been observed in a few $Ln$SbTe compounds such as CeSbTe [39], GdSbTe [26], NdSbTe [27], and TbSbTe [34], offering an approach to tune the topological states through the coupling between magnetism and topology [28,29].

In square-net systems, the presence of delocalized electrons maintains square-net lattices which is intrinsically unstable [55]. So electron counts is an effective in tuning structural and consequently the magnetic and electronic properties [40,56–63]. Substituting Te for Sb alters the electron filling and distorts the Sb square-net in $Ln$Sb$_x$Te$_{2-x}$, leading to an orthorhombic structure for off-stoichiometric $Ln$SbTe near $x = 0.70$ to $0.85$ [56–60,64]. In some $Ln$Sb$_x$Te$_{2-x}$ such as NdSb$_x$Te$_{2-x}$ [60] and PrSb$_x$Te$_{2-x}$ [40], a reemergence of tetragonal structure has been observed when Sb content $x$ is reduced below ~0.10 to 0.20. For the end compounds $Ln$Te$_2$ ($x = 0$), the tetragonal structure has been reported in CeTe$_2$ [65] and PrTe$_2$ [66], whereas in NdTe$_{1.80}$ [67] and NdTe$_{1.89}$ [68], both tetragonal and orthorhombic structures have been reported. Distorted Sb square-nets in the off-stoichiometric, orthorhombic $Ln$SbTe are shown to display charge density waves (CDWs) tunable with composition [56–58,64,69]. The evolution of CDWs results in the formation of 5-fold supercells in GdSb$_{0.48}$Te$_{1.50}$, and CeSb$_{0.51}$Te$_{1.40}$, 3-fold supercells in CeSb$_{0.3}$Te$_{1.67}$, 4-fold supercells in GdSb$_{0.21}$Te$_{1.70}$ and $3\times3\times2$ supercells in CeSb$_{0.11}$Te$_{0.89}$ [70]. CDWs have also been detected in tetragonal $Ln$SbTe [71–73] as well as in tetragonal CeTe$_2$ [74–76] and PrTe$_2$ [74], and monoclinic LaTe$_2$ [77]. The presence of CDWs significantly influences magnetism and electronic properties, leading to intricate magnetic phases [56,59,64] and diverse transport characteristics [40,58,69,78]. Recent studies on NdSb$_x$Te$_{2-x}$ revealed the existence of CDWs and the evolution of supercell structure with Sb composition [59,70] such as 5-fold supercell formation is observed in NdSb$_{0.48}$Te$_{1.37}$, ~7-fold supercells in NdSb$_{0.30}$Te$_{1.61}$, and 3-fold supercell formation in NdSb$_{0.23}$Te$_{1.70}$ [70]. The evolution of CDWs is found to couple with magnetism, resulting in changes to the magnetic structure of NdSb$_x$Te$_{2-x}$ [59,60]. A collinear in-plane antiferromagnetic (AFM) ground states in nearly stoichiometric or stoichiometric compositions change to a more complex elliptical cycloid/canted structure for intermediate Sb-compositions and finally to a collinear out-of-plane AFM structure for low Sb content in NdSb$_x$Te$_{2-x}$ [50,51].

Motivated by the complex charge and spin orders in NdSb$_x$Te$_{2-x}$, in this work we study the evolution of transport properties of this material system. Our work reveals interesting Sb-dependent resistivity behavior with Sb-Te substitution and non-monotonic evolution of MR with strong negative MR for x = 0.05 and x = 0.34, establishing this material as a promising platform to study the complex interplay of magnetism, transport, CDWs, and topological states.

## II. EXPERIMENT

Single crystals of NdSb$_x$Te$_{2-x}$ (0 < x < 1) were grown by a chemical vapor transport method with I$_2$ as the transport agent. The composition and structure of the obtained crystals were examined by x-ray diffraction and energy dispersive x-ray spectroscopy. The details of crystal growth and characterizations have been reported elsewhere [58]. Transport property measurements were performed by using a physical property measurement system (PPMS, Quantum Design).

## III. Result and discussion

Figure 1a displays the temperature dependence of resistivity for various NdSb$_x$Te$_{2-x}$. At zero magnetic field, all compositions ($0 < x < 1$) display non-metallic behavior below 300 K as manifested by resistivity increase upon cooling. While metallic transport has been observed in several $Ln$Sb$_x$Te$_{2-x}$ such as DySbTe [36] and HoSbTe [30], a similar non-metallic behavior is widely seen in many magnetic $Ln$Sb$_x$Te$_{2-x}$ such as $Ln$ = Sm [31], Ce [39], Pr [40], and Gd [52] though Fermi surface has been experimentally observed [29,31–33,51]. Such discrepancy might be associated with partial CDW gap or localization effect. As shown in Fig. 1a, for high Sb compositions ($x = 0.60$ and $0.83$) near and above the tetragonal-orthorhombic structural transitions boundary of $x \sim 0.70$ [60], the increase of resistivity with cooling is relatively weak and the resistivity values at 2 K are in the order of 0.01 Ω cm. Furthermore, for $x = 0.83$, the broad upturn near the AFM transition and the plateau-like feature is observed between 20 and 50 K likely indicate a more gradual evolution of magnetic correlations or possible magnetic frustration in this regime as discussed in Ref. [60]. In contrast, for samples with lower Sb content and well sit in the orthorhombic phase region ($0.20 < x < 0.70$), i.e., $x = 0.50, 0.44$, and $0.34$, a clear resistivity upturn occurs at low temperatures, leading to strongly enhanced resistivity value at 2 K. Specifically, for the $x = 0.44$ sample that is at the center of the orthorhombic phase region, resistivity value at 2 K is maximized to 336 Ω cm. Further reducing the Sb content to the other tetragonal-orthorhombic structural transitions boundary at $x \sim 0.20$, such as the $x = 0.24$ and $0.19$ samples, the resistivity upturn becomes less prominent. However, once the system re-enters the tetragonal phase region with even lower Sb content, sharp resistivity upturn appears again, reaching to even greater values of 3615 Ω cm and 42,888 Ω cm at 2 K for the $x = 0.10$ and $x = 0.05$ compositions, respectively. The non-monotonic evolution of electronic transport with composition in NdSb$_x$Te$_{2-x}$ might be associated with the combined effect of the composition-dependent multiple structural transitions [60], supercell structures due to CDWs [70], and magnetic structures [59,60]. Such non-monotonic evolution has also been observed in PrSb$_x$Te$_{2-x}$ which also displays re-entering of the tetragonal phase when Sb content is low [40].

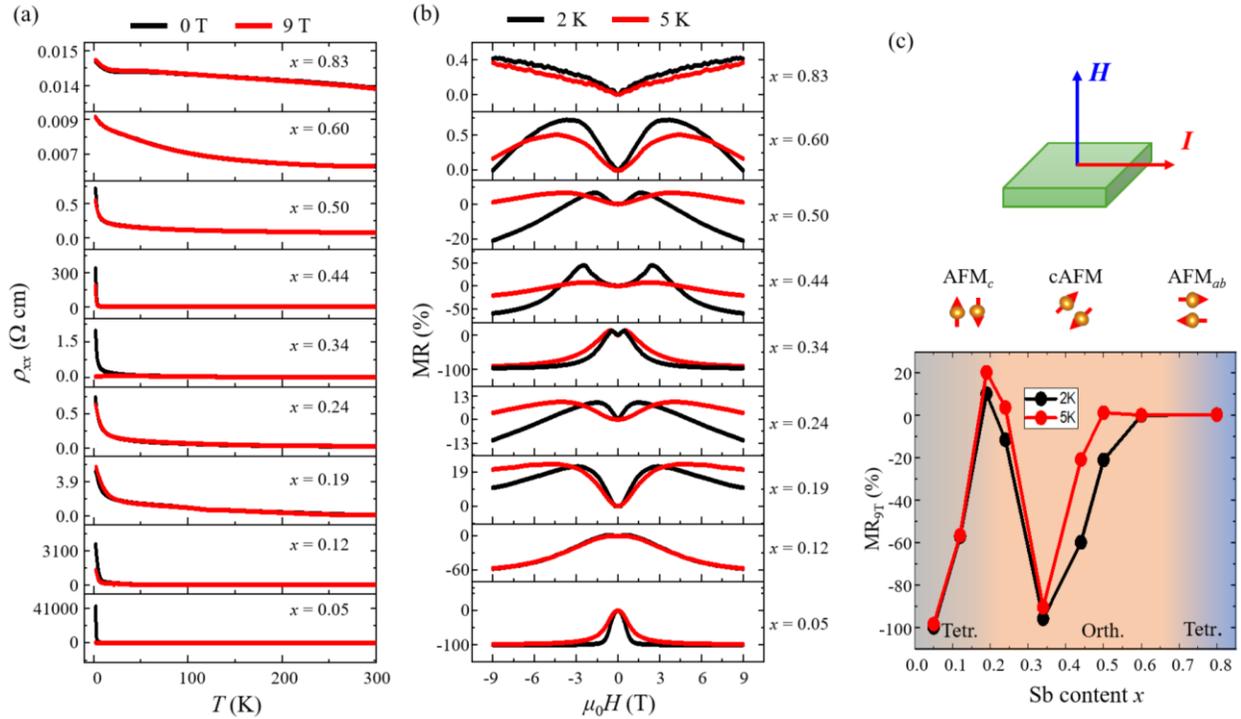

Figure 1. (a) Temperature dependent resistivity for various NdSb$_x$Te$_{2-x}$ compositions from $x = 0.05$ to $x = 0.83$ measured with magnetic field of at 0 and 9 T applied perpendicular to the *ab*-plane. (b) Normalized magnetoresistance (MR) for various NdSb$_x$Te$_{2-x}$ compositions from $x = 0.05$ to $x = 0.83$ at 2 and 5 K. (c) Evolution of 2 K and 5 K MR at 9 T with Sb content. The top schematic shows measurement configurations, as well as the moment orientations of various Sb content $x$ [60]. The AFM$_{ab}$, cAFM, and AFM$_c$ represent in-plane, canted, and out-of-plane antiferromagnetic configurations.

With the application of magnetic field along the out-of-plane direction $H//c$ (Fig. 1a), the samples near the two structure transition boundaries, such as $x = 0.83$ and 0.60, and $x = 0.24$ and

0.19, do not show remarkable modulation of in-plane resistivity $\rho_{xx}$. In contrast, a strong reduction of resistivity, especially at low temperatures, is observed for samples possessing well-defined orthorhombic (i.e., $x = 0.50$, 0.44, and 0.34) and tetragonal (i.e., $x = 0.12$ and 0.05) structures. The evolution of magnetotransport with composition is better illustrated in Fig. 1b which depicts the normalized magnetoresistivity MR $= \frac{\rho(H) - \rho(H=0)}{\rho(H=0)}$. Because the AFM ordering temperatures $T_N$ of NdSb$_x$Te$_{2-x}$ are between 2 and 3 K [27,59,60], in Fig. 1b we present MR at 2 and 5 K, i.e., below and above $T_N$. For the tetragonal, high-Sb sample ($x = 0.83$), positive MR is observed at 2 and 5 K. For orthorhombic samples in the $0.19 \leqslant x \leqslant 0.60$ composition range, MR at 2 K shows a non-monotonic field dependence, exhibiting positive MR in the low-field that is followed by a drop at higher fields. This leads to an MR dip at low field and implies coexisting and competing positive and negative MR components. A positive MR might arise from orbital MR due to Lorentz effect or weak antilocalization, etc, which is not the focus of this work. For the negative MR component, as will be shown below, we found that it is most likely driven by magnetic field-induced band modification. Additionally, the increase in the spin polarization field with Sb content in magnetization measurements [60] also aligns with the observed trend that a higher field is required for the emergence of the negative MR component in the non-monotonic MR behavior. Furthermore, such MR dip is strongly suppressed with slightly rising temperature to 5 K, i.e., just above $T_N$, implying a possible magnetic origin. A similar low-field MR dip and its suppression with increasing temperature has also been reported in PrSb$_{0.3}$Te$_{1.7}$ [40] and CeSb$_{0.11}$Te$_{1.90}$ [61]. In NdSb$_x$Te$_{2-x}$, it is worth noting that these compositions also display metamagnetic transitions at low fields, which are followed by partial moment polarization at higher fields [51]. Metamagnetic transitions and moment polarization has been reported to cause MR anomalies in many magnetic materials, such as Gd$_2$Se$_3$ [79] and EuMnBi$_2$ [80]. Hence, the complicated MR behavior in these compositions might be associated with magnetism. For samples with even lower Sb content and is thus within the tetragonal phase region, i.e., $x = 0.12$ and $0.05$, substantial negative MR is observed, as shown in Fig. 1b.

To gain a better insight into the evolution of MR with Sb content, in Fig. 1c we present the composition-dependent MR at 9 T. Overall, MR at 9 T displays similar non-monotonic composition dependences at 2 and 5 K. Above $x > 0.60$, i.e., the high-Sb tetragonal samples and samples near the tetragonal-orthorhombic phase boundary, MR at 9T is rather weak ($< 1\%$). Strong negative MR develops in orthorhombic phases by lowering the Sb content, reaching 96% in the $x = 0.34$ sample. It is worth noting that the most extensive supercell formation (~ 7 fold) due to CDWs has also been reported in a similar composition of $x = 0.30$ [70]. Further lowering the Sb content approaching the other tetragonal-orthorhombic phase boundary leads to the suppression of negative MR, and MR becomes positive around $x \sim 0.20$. For the low-Sb tetragonal sample, strong negative MR appears again, reaching the highest value of 99.94% for the $x = 0.05$ sample at 2 K and 9 T, which converts to $1.6 \times 10^5$ % if normalized to the resistivity value at 9 T. A similar complicate composition dependence for MR has also been observed in PrSb$_x$Te$_{2-x}$, which also displays two tetragonal-orthorhombic phase boundaries with negative MR maximizes for the low-Sb tetragonal and intermediate-Sb orthorhombic samples [40]. Additionally, negative MR in low-Sb ($x < 0.20$) samples appears generic for reported magnetic $Ln$Sb$_x$Te$_{2-x}$ such as PrSb$_x$Te$_{2-x}$ [40] and CeSb$_x$Te$_{2-x}$ [61], while NdSb$_{0.05}$Te$_{1.95}$ displays the strongest negative MR. Such non-monotonic evolution of resistivity, negative MR and an exceptionally large MR is reproducible in other samples with nearly identical compositions as shown in Supplemental Material [81].

Various mechanisms such as the suppression of spin scattering [79,82], chiral anomaly [12,83], spin valve effect [84], weak localization [85,86], and modification of electronic structures by magnetic fields [87,88] can give rise to substantial negative MR. Spin scattering is generally expected in magnetic systems; however, it appears to play only a minor role in NdSb$_x$Te$_{2-x}$. Spin scattering is expected to strongly influenced by magnetic phase transitions, owing to the distinct incoherent scattering from short-range order/fluctuations immediately above $T_N$ and the coherent scattering associated with long-range order below $T_N$, as observed in other materials [79]. The absence of clear transport signatures at $T_N$ therefore exclude spin scattering as the primary origin of the large negative MR. Other mechanisms listed above can also be ruled out, because chiral anomaly-induced large negative MR in TSMs occurs under parallel magnetic field and electric current configuration, spin valve effects generally require spin polarization in interlayer transport, and weak localization should display quick MR saturation at lower fields owing to the efficient suppression of quantum interference.

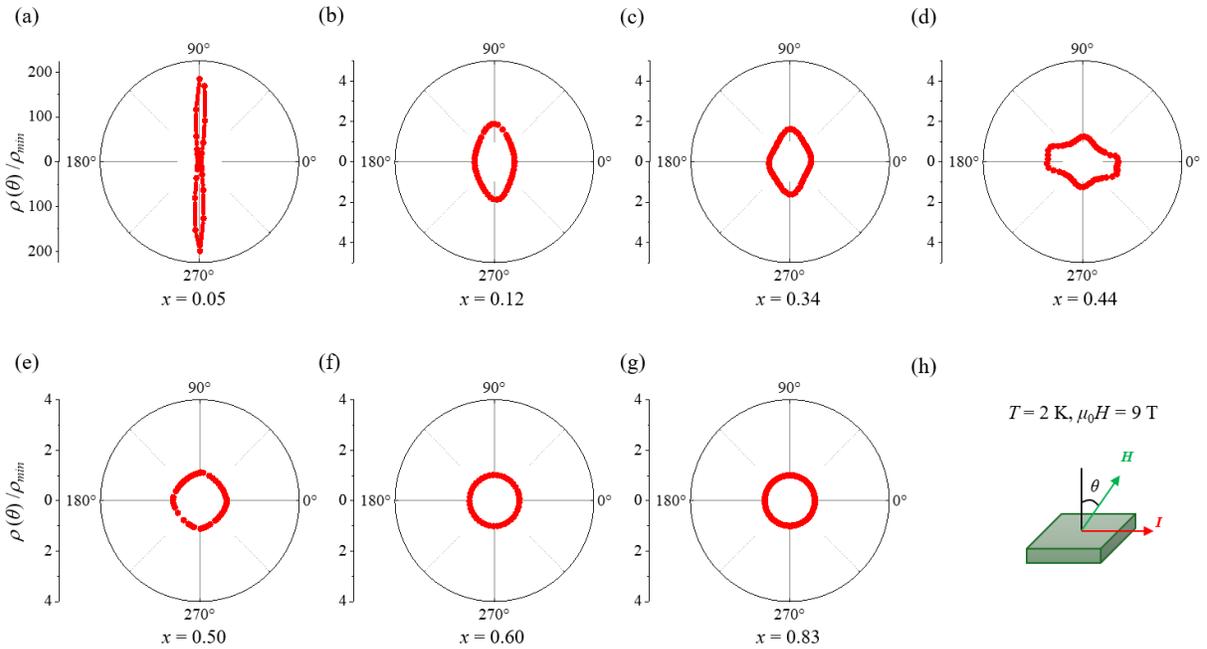

Figure 2. Polar plots of angular magnetoresistance (AMR) normalized to the resistivity minimum during the angular scan, measured under 9 T magnetic field at 2 K (below $T_N$) for NdSb$_x$Te$_{2-x}$

samples with Sb compositions $x$ = 0.05 (a), 0.12 (b), 0.34 (c), 0.44 (d), 0.50 (e), 0.60 (f), and 0.83 (g). The schematic of the measurement configuration is shown in (h).

The anisotropy of MR provides insights into its mechanism. Figures 2 and 3 displays angular dependence of magnetoresistance (AMR) for samples with various Sb contents measured with a magnetic field of 9 T rotating from the out-of-plane ($H \perp ab$, defined as $\theta = 0°$) to the in-plane ($H // ab$, defined as $\theta = 90°$) direction. Results at 2 K (Fig. 2) and 5 K (Fig. 3), which are below and above magnetic transition temperatures respectively, are presented. To facilitate comparison, the data for each sample are normalized to the minimum resistivity during the angular scan, i.e., $\rho(\theta)/\rho_{min}$. As shown in Fig. 2(a), AMR for the $x$ = 0.05 sample displays strong anisotropy with a two-fold symmetry at 2 K and 9 T, with the resistivity maximizes for $H//ab$ and minimized for $H \perp ab$. Increasing Sb content to $x$ = 0.12 retains the two-fold AMR symmetry remains, but the AMR anisotropy is reduced significantly, as shown in Fig. 2b. The AMR anisotropy continues to evolve with further increasing Sb content. The anisotropy is reduced in samples with higher Sb content, as shown in Figs. 2c-2g.

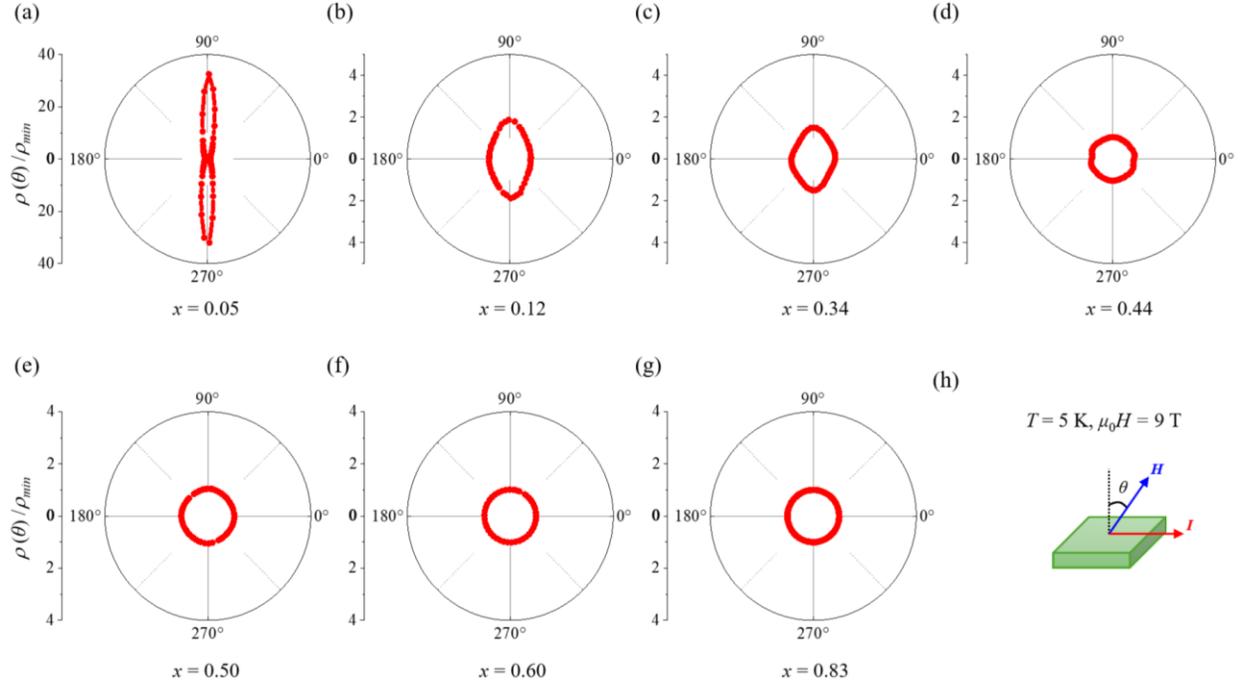

Figure 3. Polar plots of angular magnetoresistance (AMR) normalized to the resistivity minimum during the angular scan, measured under 9 T magnetic field at 5 K (above $T_N$) for NdSb$_x$Te$_{2-x}$ samples with Sb compositions $x = 0.05$ (a), 0.12 (b), 0.34 (c), 0.44 (d), 0.50 (e), 0.60 (f), and 0.83 (g). The schematic of the measurement configuration is shown in (h).

The observed anisotropy of magnetotransport in NdSb$_x$Te$_{2-x}$ is reminiscent of the anisotropy of magnetism. The low-Sb tetragonal samples with the largest negative MR for $H \perp ab$ have been reported to possess an AFM ground state with out-of-plane magnetic easy axis [60]. Increasing the Sb content above 0.20 leads to the rotation of the magnetic moment to form a cycloid magnetic structure [59] which is characterized by greatly reduced magnetic anisotropy [60]. Such correlation suggests that though magnetic scattering as a mechanism for large MR has been ruled out as stated above, the magnetotransport in NdSb$_x$Te$_{2-x}$ is correlated to magnetism. This is further supported by MR anisotropy at higher temperatures. As shown in Figs. 3a-3g, with the suppression of magnetic correlations at 5 K, i.e., immediately above $T_N$, the MR anisotropy at 9 T is also somewhat reduced. The $x = 0.44$ sample appears to show stronger $T$-dependence in MR anisotropy. This composition is close to the $x = 0.45$ which has been reported to display the highest $T_N$ [60], so that its magnetic correlations should be notably stronger and might contribute to the enhanced $T$-dependence anisotropy crossing $T_N$. In addition, for the $x = 0.83$ sample which possesses an in-plane magnetic easy axis [60], the nearly isotropic AMR appears not consistent with magnetic anisotropy. Nevertheless, it is worth noting that the MR for this sample is very small (< 1%) [Fig. 1(b)], so that the contribution to MR from magnetism is too weak to exhibit clear anisotropy.

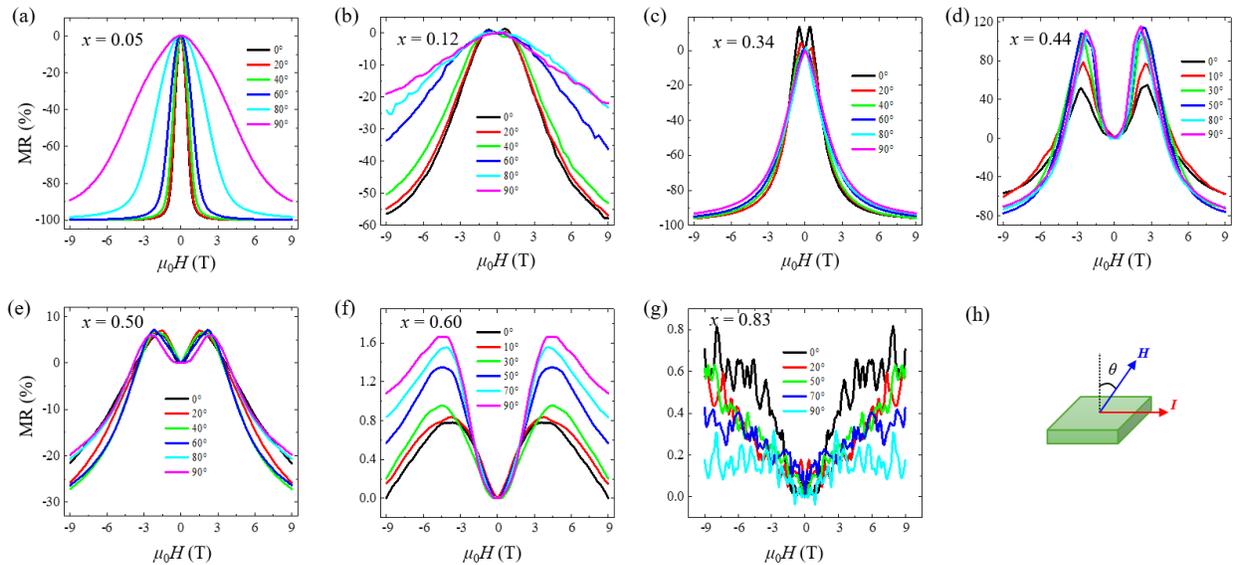

Figure 4. MR at $T = 2$ K measured under different magnetic field orientations for NdSb$_x$Te$_{2-x}$ samples with various Sb compositions $x = 0.05$ (a), 0.12 (b), 0.34 (c), 0.44 (d), 0.50 (e), 0.60 (f), 0.83 (g). The schematic of the measurement configurations is shown in (h).

In addition to the angular scan of MR at the fixed magnetic field, the evolution of MR anisotropy is also clearly manifested in the field dependence of MR measured at various field orientations. As shown in Fig. 4, the $x = 0.05$ (Fig. 4a) and 0.12 (Fig. 4b) samples display quite strong anisotropy even under low magnetic field, showing greater negative MR when field is orientated to 0° (i.e., easy axis). For intermediate Sb compositions of $x = 0.34$ (Fig. 4c), 0.44 (Fig. 4d), and 0.50 (Fig. 4e), the magnetic field orientation has weak impact on the negative MR and MR behaves more isotropic. For the $x = 0.60$ sample (Fig. 4f), the MR at 9 T appears to be anisotropic. However, such anisotropy is mainly attributed to the remarkable positive MR component. As shown in Fig. 4f, the negative MR component, which can be estimated by removing the positive MR component, is indeed rather isotropic. Therefore, the evolution of anisotropy of the negative MR across a broad composition range is in line with the reported evolution of magnetic anisotropy, i.e. out-of-plane magnetic easy axis for low-Sb compositions and canted moment for the intermediate orthorhombic compositions [60]. For samples with even higher Sb content, MR is less than 1% for the $x = 0.83$ (Fig. 4g) and 0.80 (see Supplemental Material [81]) samples and the negative MR component becomes comparable with the positive MR. Despite such weak MR, its anisotropy is in general consistent with the reported in-plane magnetic easy axis (see Supplemental Material [81]).

With the establishment of the coupling between magnetism and magnetotransport, we now discuss the mechanism for the large negative MR in certain $NdSb_xTe_{2-x}$ compositions. Since magnetic scattering as an origin has been ruled out, the large negative MR is very likely due to the impact of magnetism on the electronic band structure. The non-metallic transport for the entire composition range for $NdSb_xTe_{2-x}$ is clearly not consistent with the observed finite Fermi surface [89]. Therefore, a possible symmetry breaking or super-zone effect might gap or partially gap the system, leading to non-metallic transport. This is in line with the formation of CDW in orthorhombic $NdSb_xTe_{2-x}$ with intermediate Sb content [59,70]. A similar scenario of CDW gap has been reported in $GdSb_{0.46}Te_{1.48}$ [69] and $CeSb_{0.11}Te_{1.89}$ [61]. For low-Sb compositions that are found to display a higher symmetry tetragonal structure by $x$-ray diffraction [60], some structure distortion beyond the instrument limit might be possible as orthorhombic phase can be induced in NdTe$_2$ with Te vacancies [67,68]. Hence, modification of electronic band by magnetism to reduce band gap is a possible mechanism of strong negative MR in NdSb$_x$Te$_{2-x}$. This has been observed in other negative MR materials such as EuMnSb$_2$ [87] and GdPS [88]. Particularly, in insulating GdPS, the polarization of moment by external magnetic field leads to strong *f-d* exchange coupling that split the Gd *d*-band, which closes the band gap and induces an insulator-to-metal transition [88]. Additionally, such band engineering depends on the strength magnetization rather than the formation of the long-range magnetic order [88], so that negative MR remains remarkable above $T_N$ (Fig. 1c and Supplemental Material [81]). The effectiveness of bandgap engineering via magnetism in producing large negative MR depends on the presence of a suitably sized bandgap. If the bandgap is too large, it cannot be sufficiently reduced to enhance conductivity. Conversely, if the bandgap is too small and the system already exhibits metallic or nearly metallic transport, further reducing the bandgap has less effect on reducing resistivity. Therefore, to achieve strong negative MR, the system must possess an appropriate bandgap or partial gap that ensures semiconducting or non-metallic behavior at zero magnetic field, along with a band structure that can be significantly modified by the field to reduce the gap. In NdSb$_x$Te$_{2-x}$, the composition-dependent tetragonal-orthorhombic structural phases transition and CDW modulate the bandgap, thereby creating such "suitable" bandgap for the emergence of large negative MR in certain compositions. i.e., $x \sim 0.05$-$0.10$ and $0.30$-$0.45$.

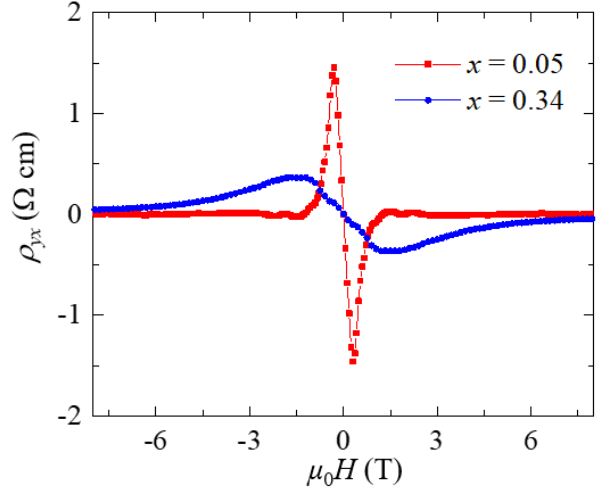

Figure 5. Hall effect at $T = 2$ K for the $x = 0.05$ and 0.34 samples.

The magnetic field-induced band engineering mechanism is supported by the Hall effect measurements on the two compositions $x = 0.05$ and 0.34 exhibiting the strongest negative MR. As shown in Fig. 5, both samples display clear anomalous Hall features that emerge in the same magnetic field range where negative MR becomes prominent. For the $x = 0.05$ sample, the anomalous Hall signal appears at low fields, consistent with the sharp onset of strong negative MR at low field (Fig. 1b). In contrast, the $x = 0.34$ sample exhibits a broader anomalous Hall feature that aligns well with the more gradual field dependence of its MR (Fig. 1b). This behavior is also reported in related material $CeSb_xTe_{2-x}$ [61], where anomalous Hall responses have also been linked to field-induced band modifications.

In conclusion, we observed the non-monotonic evolution of MR with Sb compositions in $NdSb_xTe_{2-x}$ and strongest negative MR in $x = 0.05$ and 0.34 compositions. The strong magnetoresistance in this material establishes it as a promising platform for investigating and engineering topological semimetal for future device applications. The correlation between magnetic and transport anisotropy reveals the interplay between them. Together with the structure evolution with composition, it is highly likely that the band modification due to the magnetic field leads to large negative MR in NdSb$_x$Te$_{2-x}$. Importantly, such a mechanism might be generic in various $Ln$Sb$_x$Te$_{2-x}$ compounds given their similarities in structure, magnetism, and transport properties.

**Acknowledgement**

This work was primarily supported by the U.S. Department of Energy (DOE), Office of Science, Basic Energy Sciences program under Grant No. DE-SC0022006 (crystal growth and magnetotransport). We acknowledge the MonArk NSF Quantum Foundry for resistivity sample fabrication, which is supported by the National Science Foundation (NSF) Q-AMASE-i program under NSF Award No. DMR-1906383.

**References**

[1]   B. Yan and C. Felser, Topological Materials: Weyl Semimetals, Annu. Rev. Condens. Matter Phys. **8**, 337 (2017).

[2]   N. P. Armitage, E. J. Mele, and A. Vishwanath, Weyl and Dirac semimetals in three-dimensional solids, Rev. Mod. Phys. **90**, 015001 (2018).

[3]   A. Bernevig, H. Weng, Z. Fang, and X. Dai, Recent Progress in the Study of Topological Semimetals, J. Phys. Soc. Jpn. **87**, 041001 (2018).

[4]   J. Hu, S.-Y. Xu, N. Ni, and Z. Mao, Transport of Topological Semimetals, Annu. Rev. Mater. Res. **49**, 207 (2019).

[5] S.-Y. Xu, C. Liu, S. K. Kushwaha, R. Sankar, J. W. Krizan, I. Belopolski, M. Neupane, G. Bian, N. Alidoust, T.-R. Chang et al., Observation of Fermi arc surface states in a topological metal, Science **347**, 294 (2015).

[6] S.-Y. Xu, N. Alidoust, I. Belopolski, Z. Yuan, G. Bian, T.-R. Chang, H. Zheng, V. N. Strocov, D. S. Sanchez, G. Chang et al., Discovery of a Weyl fermion state with Fermi arcs in niobium arsenide, Nat. Phys. **11**, 748 (2015).

[7] S.-M. Huang, S.-Y. Xu, I. Belopolski, C.-C. Lee, G. Chang, B. Wang, N. Alidoust, G. Bian, M. Neupane, C. Zhang et al., A Weyl Fermion semimetal with surface Fermi arcs in the transition metal monopnictide TaAs class, Nat. Commun. **6**, 7373 (2015).

[8] T. Liang, Q. Gibson, M. N. Ali, M. Liu, R. J. Cava, and N. P. Ong, Ultrahigh mobility and giant magnetoresistance in the Dirac semimetal $Cd_3As_2$, Nat. Mater. **14**, 3 (2015).

[9] M. Neupane, S.-Y. Xu, R. Sankar, N. Alidoust, G. Bian, C. Liu, I. Belopolski, T.-R. Chang, H.-T. Jeng, H. Lin, A. Bansil, F. Chou & M. Z. Hasan, Observation of a three-dimensional topological Dirac semimetal phase in high-mobility $Cd_3As_2$, Nat. Commun. **5**, 3786 (2014).

[10] C. Shekhar, A. K. Nayak, Y. Sun, M. Schmidt, M. Nicklas, I. Leermakers, U. Zeitler, Y. Skourski, J. Wosnitza, Z. Liu et al., Extremely large magnetoresistance and ultrahigh mobility in the topological Weyl semimetal candidate NbP, Nat. Phys. **11**, 645 (2015).

[11] A. A. Burkov, Chiral anomaly and transport in Weyl metals, J. Phys. Condens. Matter **27**, 113201 (2015).

[12] Q. Li, D. E. Kharzeev, C. Zhang, Y. Huang, I. Pletikosić, A. V. Fedorov, R. D. Zhong, J. A. Schneeloch, G. D. Gu, and T. Valla, Chiral magnetic effect in $ZrTe_5$, Nat. Phys. **12**, 6 (2016).

[13]     E. Li, Y. Sun, N. Kumar, L. Muechler, A. Sun, L. Jiao, S.-Y. Yang, D. Liu, A. Liang, Q. Xu et al., Giant anomalous Hall effect in a ferromagnetic kagome-lattice semimetal, Nat. Phys. **14**, 1125 (2018).

[14]     S. N. Guin, P. Vir, Y. Zhang, N. Kumar, S. J. Watzman, C. Fu, E. Liu, K. Manna, W. Schnelle, J. Gooth et al., Zero-Field Nernst Effect in a Ferromagnetic Kagome-Lattice Weyl-Semimetal $Co_3Sn_2S_2$, Adv. Mater. **31**, 1806622 (2019).

[15]     M. Kang, L. Ye, S. Fang, J.-S. You, A. Levitan, M. Han, J. I. Facio, C. Jozwiak, A. Bostwick, E. Rotenberg et al., Dirac fermions and flat bands in the ideal kagome metal FeSn, Nat. Mater. **19**, 163 (2020).

[16]     K. Kim, J. Seo, E. Lee, K.-T. Ko, B. S. Kim, B. G. Jang, J. M. Ok, J. Lee, Y. J. Jo, W. Kang, et al., Large anomalous Hall current induced by topological nodal lines in a ferromagnetic van der Waals semimetal, Nat. Mater. **17**, 794 (2018).

[17]     S. Nakatsuji, N. Kiyohara, and T. Higo, Large anomalous Hall effect in a non-collinear antiferromagnet at room temperature, Nature **527**, 212 (2015).

[18]     M. Ikhlas, T. Tomita, T. Koretsune, M.-T. Suzuki, D. Nishio-Hamane, R. Arita, Y. Otani, and S. Nakatsuji, Large anomalous Nernst effect at room temperature in a chiral antiferromagnet, Nat. Phys. **13**, 1085 (2017).

[19]     A. K. Nayak, J. E. Fischer, Y. Sun, B. Yan, J. Karel, A. C. Komarek, C. Shekhar, N. Kumar, W. Schnelle, J. Kübler, C. Felser, and S. S. P. Parkin, Large anomalous Hall effect driven by a nonvanishing Berry curvature in the noncolinear antiferromagnet $Mn_3Ge$, Sci. Adv. **2**, e1501870 (2016).

[20]   C. Wuttke, F. Caglieris, S. Sykora, F. Scaravaggi, A. U. B. Wolter, K. Manna, V. Süss, C. Shekhar, C. Felser, B. Büchner, and C. Hess, Berry curvature unravelled by the anomalous Nernst effect in $Mn_3Ge$, Phys. Rev. B **100**, 085111 (2019).

[21]   J. Noky, Y. Zhang, J. Gooth, C. Felser, and Y. Sun, Giant anomalous Hall and Nernst effect in magnetic cubic Heusler compounds, Npj Comput. Mater. **6**, 77 (2020).

[22]   P. Li, J. Koo, W. Ning, J. Li, L. Miao, L. Min, Y. Zhu, Y. Wang, N. Alem, C.-X. Liu, Z. Mao and B. Yan, Giant room temperature anomalous Hall effect and tunable topology in a ferromagnetic topological semimetal $Co_2MnAl$, Nat. Commun. **11**, 3476 (2020).

[23]   A. Sakai, Y. P. Mizuta, A. A. Nugroho, R. Sihombing, T. Koretsune, M.-T. Suzuki, N. Takemori, R. Ishii, D. Nishio-Hamane, R. Arita, P. Goswami and S. Nakatsuji, Giant anomalous Nernst effect and quantum-critical scaling in a ferromagnetic semimetal, Nat. Phys. **14**, 1119 (2018).

[24]   K. Iwaya, Y. Kohsaka, K. Okawa, T. Machida, M. S. Bahramy, T. Hanaguri, and T. Sasagawa, Full-gap superconductivity in spin-polarised surface states of topological semimetal β-$PdBi_2$, Nat. Commun. **8**, 976 (2017).

[25]   B. Ghosh, D. Mondal, C.-N. Kuo, C. S. Lue, J. Nayak, J. Fujii, I. Vobornik, A. Politano, and A. Agarwal, Observation of bulk states and spin-polarized topological surface states in transition metal dichalcogenide Dirac semimetal candidate $NiTe_2$, Phys. Rev. B **100**, 195134 (2019).

[26]   M. Neupane, N. Alidoust, M. M. Hosen, J.-X. Zhu, K. Dimitri, S.-Y. Xu, N. Dhakal, R. Sankar, I. Belopolski, D. S. Sanchez et al., Observation of the spin-polarized surface state in a noncentrosymmetric superconductor BiPd, Nat. Commun. **7**, 13315 (2016).

[27] K. Pandey, R. Basnet, A. Wegner, G. Acharya, M. R. U. Nabi, J. Liu, J. Wang, Y. K. Takahashi, B. Da, and J. Hu, Electronic and magnetic properties of the topological semimetal candidate NdSbTe, Phys. Rev. B **101**, 235161 (2020).

[28] M. M. Hosen, G. Dhakal, K. Dimitri, P. Maldonado, A. Aperis, F. Kabir, C. Sims, P. Riseborough, P. M. Oppeneer, D. Kaczorowski, T. Durakiewicz and M. Neupane, Discovery of topological nodal-line fermionic phase in a magnetic material GdSbTe, Sci. Rep. **8**, 13283 (2018).

[29] L. M. Schoop, A. Topp, J. Lippmann, F. Orlandi, L. Müchler, M. G. Vergniory, Y. Sun, A. W. Rost, V. Duppel, M. Krivenkov et al., Tunable Weyl and Dirac states in the nonsymmorphic compound CeSbTe, Sci. Adv. **4**, eaar2317 (2018).

[30] M. Yang, Y. Qian, D. Yan, Y. Li, Y. Song, Z. Wang, C. Yi, H. L. Feng, H. Weng, and Y. Shi, Magnetic and electronic properties of a topological nodal line semimetal candidate: HoSbTe, Phys. Rev. Mater. **4**, 094203 (2020).

[31] K. Pandey, D. Mondal, J. W. Villanova, J. Roll, R. Basnet, A. Wegner, G. Acharya, M. R. U. Nabi, B. Ghosh, J. Fujii, J. Wang et al., Magnetic Topological Semimetal Phase with Electronic Correlation Enhancement in SmSbTe, Adv. Quantum Technol. **4**, 2100063 (2021).

[32] D. Yuan, D. Huang, X. Ma, X. Chen, H. Ren, Y. Zhang, W. Feng, X. Zhu, B. Wang, X. He et al., Observation of Dirac nodal line states in topological semimetal candidate PrSbTe, Phys. Rev. B **109**, 045113 (2024).

[33] S. Regmi, I. B. Elius, A. P. Sakhya, M. Sprague, M. I. Mondal, N. Valadez, V. Buturlim, K. Booth, T. Romanova, K. Gofryk et al., Electronic structure in a rare-earth based nodal-line semimetal candidate PrSbTe, Phys. Rev. Mater. **8**, L041201 (2024).

[34] F. Gao, J. Huang, W. Ren, H. Wu, M. An, X. Wu, L. Zhang, T. Yang, A. Wang, Y. Chai, X. Zhao, T. Yang, B. Li, and Z. Zhang, Magnetic and Magnetotransport Properties of the Magnetic Topological Nodal-Line Semimetal TbSbTe, Adv. Quantum Technol. **6**, 2200163 (2023).

[35] I. Plokhikh, V. Pomjakushin, D. J. Gawryluk, O. Zaharko, and E. Pomjakushina, Competing Magnetic Phases in *Ln*SbTe ( *Ln* = Ho and Tb), Inorg. Chem. **61**, 11399 (2022).

[36] F. Gao, J. Huang, W. Ren, M. Li, H. Wang, T. Yang, B. Li, and Z. Zhang, Magnetic and transport properties of the topological compound DySbTe, Phys. Rev. B **105**, 214434 (2022).

[37] I. Plokhikh, V. Pomjakushin, D. Jakub Gawryluk, O. Zaharko, and E. Pomjakushina, On the magnetic structures of 1:1:1 stoichiometric topological phases LnSbTe (Ln = Pr, Nd, Dy and Er), J. Magn. Magn. Mater. **583**, 171009 (2023).

[38] S. Yue, Y. Qian, M. Yang, D. Geng, C. Yi, S. Kumar, K. Shimada, P. Cheng, L. Chen, Z. Wang et al., Topological electronic structure in the antiferromagnet HoSbTe, Phys. Rev. B **102**, 155109 (2020).

[39] B. Lv, J. Chen, L. Qiao, J. Ma, X. Yang, M. Li, M. Wang, Q. Tao, and Z.-A. Xu, Magnetic and transport properties of low-carrier-density Kondo semimetal CeSbTe, J. Phys. Condens. Matter **31**, 355601 (2019).

[40] G. Acharya, K. Pandey, M. M. Sharma, J. Wang, S. Karki Chhetri, M. R. U. Nabi, D. Upreti, R. Basnet, J. Sakon, and J. Hu, Large negative magnetoresistance in the off-stoichiometric topological semimetal PrSb$_x$Te$_{2-x}$, Phys. Rev. B **111**, 024421 (2025).

[41] Q. Xu, Z. Song, S. Nie, H. Weng, Z. Fang, and X. Dai, Two-dimensional oxide topological insulator with iron-pnictide superconductor LiFeAs structure, Phys. Rev. B **92**, 205310 (2015).

[42] L. M. Schoop, M. N. Ali, C. Straßer, A. Topp, A. Varykhalov, D. Marchenko, V. Duppel, S. S. P. Parkin, B. V. Lotsch, and C. R. Ast, Dirac cone protected by non-symmorphic symmetry and three-dimensional Dirac line node in ZrSiS, Nat. Commun. **7**, 11696 (2016).

[43] J. Hu, Z. Tang, J. Liu, X. Liu, Y. Zhu, D. Graf, K. Myhro, S. Tran, C. N. Lau, J. Wei and Z. Mao, Evidence of Topological Nodal-Line Fermions in ZrSiSe and ZrSiTe, Phys. Rev. Lett. **117**, 016602 (2016).

[44] M. Neupane, I. Belopolski, M. M. Hosen, D. S. Sanchez, R. Sankar, M. Szlawska, S.-Y. Xu, K. Dimitri, N. Dhakal, Pablo Maldonado et al., Observation of topological nodal fermion semimetal phase in ZrSiS, Phys. Rev. B **93**, 201104 (2016).

[45] Y. Shao, A. N. Rudenko, J. Hu, Z. Sun, Y. Zhu, S. Moon, A. J. Millis, S. Yuan, A. I. Lichtenstein, D. Smirnov, Z. Q. Mao, M. I. Katsnelson and D. N. Basov, Electronic correlations in nodal-line semimetals, Nat. Phys. **16**, 636 (2020).

[46] C. C. Gu, J. Hu, X. L. Chen, Z. P. Guo, B. T. Fu, Y. H. Zhou, C. An, Y. Zhou, R. R. Zhang, C. Y. Xi et al., Experimental evidence of crystal symmetry protection for the topological nodal line semimetal state in ZrSiS, Phys. Rev. B **100**, 205124 (2019).

[47] D. VanGennep, T. A. Paul, C. W. Yerger, S. T. Weir, Y. K. Vohra, and J. J. Hamlin, Possible pressure-induced topological quantum phase transition in the nodal line semimetal ZrSiS, Phys. Rev. B **99**, 085204 (2019).

[48] A. Topp, R. Queiroz, A. Grüneis, L. Müchler, A. W. Rost, A. Varykhalov, D. Marchenko, M. Krivenkov, F. Rodolakis, J. L. McChesney et al., Surface Floating 2D Bands in Layered Nonsymmorphic Semimetals: ZrSiS and Related Compounds, Phys. Rev. X **7**, 041073 (2017).

[49] X. Liu, C. Yue, S. V. Erohin, Y. Zhu, A. Joshy, J. Liu, A. M. Sanchez, D. Graf, P. B. Sorokin, Z. Mao, J. Hu and J. Wei, Quantum Transport of the 2D Surface State in a Nonsymmorphic Semimetal, Nano Lett. **21**, 4887 (2021).

[50] R. Sankar, I. P. Muthuselvam, K. Rajagopal, K. Ramesh Babu, G. S. Murugan, K. S. Bayikadi, K. Moovendaran, C. Ting Wu, and G.-Y. Guo, Anisotropic Magnetic Properties of Nonsymmorphic Semimetallic Single Crystal NdSbTe, Cryst. Growth Des. **20**, 6585 (2020).

[51] S. Regmi, G. Dhakal, F. C. Kabeer, N. Harrison, F. Kabir, A. P. Sakhya, K. Gofryk, D. Kaczorowski, P. M. Oppeneer, and M. Neupane, Observation of multiple nodal lines in SmSbTe, Phys. Rev. Mater. **6**, L031201 (2022).

[52] R. Sankar, I. P. Muthuselvam, K. R. Babu, G. S. Murugan, K. Rajagopal, R. Kumar, T.-C. Wu, C.-Y. Wen, W.-L. Lee, G.-Y. Guo, and F.-C. Chou, Crystal Growth and Magnetic Properties of Topological Nodal-Line Semimetal GdSbTe with Antiferromagnetic Spin Ordering, Inorg. Chem. **58**, 11730 (2019).

[53] R. Singha, A. Pariari, B. Satpati, and P. Mandal, Magnetotransport properties and evidence of a topological insulating state in LaSbTe, Phys. Rev. B **96**, 245138 (2017).

[54] K. Pandey, L. Sayler, R. Basnet, J. Sakon, F. Wang, and J. Hu, Crystal Growth and Electronic Properties of LaSbSe, Crystals **12**, 1663 (2022).

[55] S. Klemenz, A. K. Hay, S. M. L. Teicher, A. Topp, J. Cano, and L. M. Schoop, The Role of Delocalized Chemical Bonding in Square-Net-Based Topological Semimetals, J. Am. Chem. Soc. **142**, 6350 (2020).

[56] S. Lei, A. Saltzman, and L. M. Schoop, Complex magnetic phases enriched by charge density waves in the topological semimetals GdSb$_x$Te$_{2-x-\delta}$, Phys. Rev. B **103**, 134418 (2021).

[57] S. Lei, V. Duppel, J. M. Lippmann, J. Nuss, B. V. Lotsch, and L. M. Schoop, Charge Density Waves and Magnetism in Topological Semimetal Candidates GdSb$_x$Te$_{2-x-\delta}$, Adv. Quantum Technol. **2**, 1900045 (2019).

[58] K. Pandey, R. Basnet, J. Wang, B. Da, and J. Hu, Evolution of electronic and magnetic properties in the topological semimetal SmSb$_x$Te$_{2-x}$, Phys. Rev. B **105**, 155139 (2022).

[59] T. H. Salters, F. Orlandi, T. Berry, J. F. Khoury, E. Whittaker, P. Manuel, and L. M. Schoop, Charge density wave-templated spin cycloid in topological semimetal NdSb$_x$Te$_{2-x-\delta}$, Phys. Rev. Mater. **7**, 044203 (2023).

[60] S. Karki Chhetri, R. Basnet, J. Wang, K. Pandey, G. Acharya, M. R. U. Nabi, D. Upreti, J. Sakon, M. Mortazavi, and J. Hu, Evolution of magnetism in the magnetic topological semimetal NdSb$_x$Te$_{2-x+\delta}$, Phys. Rev. B **109**, 184429 (2024).

[61] R. Singha, K. J. Dalgaard, D. Marchenko, M. Krivenkov, E. D. L. Rienks, M. Jovanovic, S. M. L. Teicher, J. Hu, T. H. Salters, J. Lin et al., Colossal magnetoresistance in the multiple wave vector charge density wave regime of an antiferromagnetic Dirac semimetal, Sci. Adv. **9**, eadh0145 (2023).

[62] J. Bannies, M. Michiardi, H.-H. Kung, S. Godin, J. W. Simonson, M. Oudah, M. Zonno, S. Gorovikov, S. Zhdanovich, I. S. Elfimov, A. Damascelli, M. C. Aronson, *Electronically-Driven Switching of Topology in LaSbTe*, https://arxiv.org/abs/2407.08798

[63]    E. DiMasi, B. Foran, M. C. Aronson, and S. Lee, Stability of charge-density waves under continuous variation of band filling in LaTe$_{2-x}$Sb$_x$ ($0 \leqslant x \leqslant 1$), Phys. Rev. B **54**, 13587 (1996).

[64]    R. Singha, T. H. Salters, S. M. L. Teicher, S. Lei, J. F. Khoury, N. P. Ong, and L. M. Schoop, Evolving Devil's Staircase Magnetization from Tunable Charge Density Waves in Nonsymmorphic Dirac Semimetals, Adv. Mater. **33**, 2103476 (2021).

[65]    K. Stöwe, Crystal structure and magnetic properties of CeTe$_2$, J. Alloys Compd. **307**, 101 (2000).

[66]    K. Stöwe, Die Kristallstruktur von PrTe$_2$, Z. Für Anorg. Allg. Chem. **626**, 803 (2000).

[67]    R. Wang, H. Steinfink, and W. F. Bradley, The Crystal Structure of Lanthanum Telluride and of Tellurium-Deficient Neodymium Telluride, Inorg. Chem. **5**, 142 (1966).

[68]    K. Stöwe, Crystal structure, magnetic properties and band gap measurements of NdTe$_{2-x}$ ($x$=0.11(1)), Z. Für Krist. - Cryst. Mater. **216**, 215 (2001).

[69]    S. Lei, S. M. L. Teicher, A. Topp, K. Cai, J. Lin, G. Cheng, T. H. Salters, F. Rodolakis, J. L. McChesney, S. Lapidus et al., Band Engineering of Dirac Semimetals Using Charge Density Waves, Adv. Mater. **33**, 2101591 (2021).

[70]    T. H. Salters, J. Colagiuri, A. Koch Liston, J. Leeman, T. Berry, and L. M. Schoop, Synthesis and Stability Phase Diagram of Topological Semimetal Family LnSb$_x$Te$_{2-x-\delta}$, Chem. Mater. **36**, 11873 (2024).

[71]    P. Li, B.J. Lv, Y. Fang, W. Guo, Z.Z. Wu, Y. Wu, D.W. Shen, Y.F. Nie, L. Petaccia, C. Cao, Z.-A. Xu and Y. Liu Charge density wave and weak Kondo effect in a Dirac semimetal CeSbTe, Sci. China Phys. Mech. Astron. **64**, 237412 (2021).

[72] L. Cao, C. Zhang, Y. Yang, L. Wang, B. Gao, X. Wang, Y. Shi, and R. Chen, *Multiple Charge-Density-Wave Gaps in LaSbTe and CeSbTe as Revealed by Ultrafast Spectroscopy*.

[73] L. Y. Cao, M. Yang, L. Wang, Y. Li, B. X. Gao, L. Wang, J. L. Liu, A. F. Fang, Y. G. Shi, and R. Y. Chen, Optical study of the topological materials LnSbTe ( Ln = La , Ce , Sm , Gd ), Phys. Rev. B **106**, 245145 (2022).

[74] E. Lee, D. H. Kim, J. D. Denlinger, J. Kim, K. Kim, B. I. Min, B. H. Min, Y. S. Kwon, and J.-S. Kang, Angle-resolved and resonant photoemission spectroscopy study of the Fermi surface reconstruction in the charge density wave systems $CeTe_2$ and $PrTe_2$, Phys. Rev. B **91**, 125137 (2015).

[75] J.-S. Kang, D. H. Kim, H. J. Lee, J. Hwang, H.-K. Lee, H.-D. Kim, B. H. Min, K. E. Lee, Y. S. Kwon, and J. W. Kim, Fermi surface reconstruction in $CeTe_2$ induced by charge density waves investigated via angle resolved photoemission, Phys. Rev. B **85**, 085104 (2012).

[76] K. Y. Shin, V. Brouet, N. Ru, Z. X. Shen, and I. R. Fisher, Electronic structure and charge-density wave formation in $LaTe_{1.95}$ and $CeTe_{2.00}$, Phys. Rev. B **72**, 085132 (2005).

[77] D. R. Garcia, G.-H. Gweon, S. Y. Zhou, J. Graf, C. M. Jozwiak, M. H. Jung, Y. S. Kwon, and A. Lanzara, Revealing Charge Density Wave Formation in the $LaTe_2$ System by Angle Resolved Photoemission Spectroscopy, Phys. Rev. Lett. **98**, 166403 (2007).

[78] H. Murakawa, Y. Nakaoka, K. Iwase, T. Kida, M. Hagiwara, H. Sakai, and N. Hanasaki, Giant negative magnetoresistance in the layered semiconductor $CeTe_{2-x}Sb_x$ with variable magnetic polaron density, Phys. Rev. B **107**, 165138 (2023).

[79] S. Karki Chhetri, G. Acharya, D. Graf, R. Basnet, S. Rahman, M. M. Sharma, D. Upreti, M. R. U. Nabi, S. Kryvyi, Josh Sakon et al., Large negative magnetoresistance in antiferromagnetic $Gd_2Se_3$, Phys. Rev. B **111**, 014431 (2025).

[80] A. F. May, M. A. McGuire, and B. C. Sales, Effect of Eu magnetism on the electronic properties of the candidate Dirac material $EuMnBi_2$, Phys. Rev. B **90**, 075109 (2014).

[81] See Supplemental Material at http://link.aps.org/supplemental/10.1103/lh2y-nws6 for additional magnetotransport data collected from different $NdSb_xTe_{2-x}$ single crystals samples.

[82] A. M. Goforth, P. Klavins, J. C. Fettinger, and S. M. Kauzlarich, Magnetic Properties and Negative Colossal Magnetoresistance of the Rare Earth Zintl phase $EuIn_2As_2$, Inorg. Chem. **47**, 11048 (2008).

[83] J. Xiong, S. K. Kushwaha, T. Liang, J. W. Krizan, M. Hirschberger, W. Wang, R. J. Cava, and N. P. Ong, Evidence for the chiral anomaly in the Dirac semimetal $Na_3Bi$, Science **350**, 413 (2015).

[84] P. A. Grünberg, Nobel Lecture: From spin waves to giant magnetoresistance and beyond, Rev. Mod. Phys. **80**, 1531 (2008).

[85] G. Bergmann, Weak localization in thin films, Phys. Rep. **107**, 1 (1984).

[86] J. J. Lin and J. P. Bird, Recent experimental studies of electron dephasing in metal and semiconductor mesoscopic structures, J. Phys. Condens. Matter **14**, R501 (2002).

[87] Z. L. Sun, A. F. Wang, H. M. Mu, H. H. Wang, Z. F. Wang, T. Wu, Z. Y. Wang, X. Y. Zhou, and X. H. Chen, Field-induced metal-to-insulator transition and colossal anisotropic magnetoresistance in a nearly Dirac material $EuMnSb_2$, Npj Quantum Mater. **6**, 94 (2021).

[88]     G. Acharya, B. Neupane, C.-H. Hsu, X. P. Yang, D. Graf, E. S. Choi, K. Pandey, M. R. U. Nabi, S. Karki Chhetri, R. Basnet et al., Insulator-to-Metal Transition and Isotropic Gigantic Magnetoresistance in Layered Magnetic Semiconductors, Adv. Mater. **36**, 2410655 (2024).

[89]     S. Regmi, R. Smith, A. P. Sakhya, M. Sprague, M. I. Mondal, I. B. Elius, N. Valadez, A. Ptok, D. Kaczorowski, and M. Neupane, Observation of gapless nodal-line states in NdSbTe, Phys. Rev. Mater. **7**, 044202 (2023).